\documentclass[11pt]{article}%
\usepackage{amsmath}
\usepackage{amsfonts}
\usepackage{amssymb}
\usepackage[latin1]{inputenc}
\usepackage{graphicx}
\usepackage{epstopdf}
\usepackage{graphicx}
\usepackage{subfigure}%
\usepackage{natbib}
\usepackage{RR}

\providecommand{\U}[1]{\protect\rule{.1in}{.1in}}

\graphicspath{{./}{./figures/}{./logos/}}

\RRdate{November 2010}

\RRauthor{
Fran\c cois Caron\thanks{INRIA Bordeaux Sud-Ouest \& Institut de Mathématiques de Bordeaux. Universit\'{e} Bordeaux I, 33405 Talence cedex, France. \texttt{Francois.Caron@inria.fr}}%
  \and
Arnaud Doucet\thanks{Departments of Computer Science and Statistics, University of British Columbia, Vancouver, BC, V6T 1Z2, Canada. \texttt{Arnaud@stat.ubc.ca}}%
}
\authorhead{F. Caron \& A. Doucet}
\RRtitle{Inférence bayésienne pour les modèles de Bradley-Terry généralisés}
\RRetitle{Efficient Bayesian Inference for Generalized Bradley-Terry Models}
\titlehead{Bayesian Inference for Bradley-Terry Models}
\RRresume{Le modèle de Bradley-Terry est une approche populaire pour décrire les résultats possibles lorsque des éléments d'un ensemble sont mis en comparaison par paire. Il a trouvé de nombreuses applications incluant le comportement animal, le classement de joueurs d'échecs et la classification multi-classes. Plusieurs extensions du modèle classique ont été proposées dans la littérature afin de prendre en compte des matchs nuls, des comparaisons multiples et des comparaisons entre groupes. D'un point de vue computationel, \cite{Hunter2004} a proposé des algorithmes MM (minorization-maximization) itératifs efficaces pour l'estimation du maximum de vraisemblance dans les modèles de Bradley-Terry généralisés, tandis que l'inférence bayésienne est réalisée à l'aide d'algorithmes MCMC (Markov chain Monte Carlo) basées sur des lois de proposition Metropolis-Hastings (M-H) adaptées. Nous montrons que ces algorithmes MM peuvent être réinterprétés comme des instances d'algorithmes EM (Expectation-Maximization) associés à des ensembles de variables latentes. Ces variables latentes nous permettent de dériver des algorithmes de Gibbs simples pour l'inférence bayésiennes. Nous démontrons expérimentalement l'efficacité de ces algorithmes sur plusieurs applications.
}
\RRabstract{The Bradley-Terry model is a popular approach to describe probabilities of the
possible outcomes when elements of a set are repeatedly compared with one
another in pairs. It has found many applications including animal behaviour,
chess ranking and multiclass classification. Numerous extensions of the basic
model have also been proposed in the literature including models with ties,
multiple comparisons, group comparisons and random graphs. From a
computational point of view, \cite{Hunter2004} has proposed efficient
iterative MM\ (minorization-maximization) algorithms to perform maximum
likelihood estimation for these generalized Bradley-Terry models whereas
Bayesian inference is typically performed using MCMC\ (Markov chain Monte
Carlo) algorithms based on tailored Metropolis-Hastings (M-H) proposals. We
show here that these MM\ algorithms can be reinterpreted as special instances
of Expectation-Maximization (EM) algorithms associated to suitable sets of
latent variables and propose some original extensions. These latent variables
allow us to derive simple Gibbs samplers for Bayesian inference. We
demonstrate experimentally the efficiency of these algorithms on a variety of applications.}
\RRmotcle{Modèle de Bradley-Terry, Algorithme EM, Echantillonneur de Gibbs, Estimation par maximum de vraisemblance, algorithmes MCMC, algorithme MM, Modèle de Plackett-Luce}
\RRkeyword{Bradley--Terry model, data augmentation, EM algorithm, Gibbs
sampler, maximum likelihood estimation, MCMC algorithms, MM algorithm,
Plackett--Luce model}
 \RRprojet{ALEA}  
\RRdomaine{1} 
\RRtheme{Stochastic Methods and Models}
\RCBordeaux 

\begin{document}

\RRNo{7445}

\makeRR   %

\section{Introduction}

Consider\ a set of $K$ elements. These elements are repeatedly compared with
one another in pairs. For two elements $i$ and $j$ of this set, \cite{Bradley1952} suggested the following model%
\begin{equation}
\Pr(i\text{ beats }j)=\frac{\lambda_{i}}{\lambda_{i}+\lambda_{j}}
\label{eq:BTmodel}%
\end{equation}
where $\lambda_{l}>0$ is a parameter associated to element $l\in\left\{
1,2,\ldots,K\right\}  $ that represents its skill rating and we denote
$\lambda:=\left\{  \lambda_{i}\right\}  _{i=1}^{K}$.

This model has found numerous applications. As mentioned in \citep{Hunter2004},
as early as 1976, a published bibliography on paired comparisons includes
several hundred entries~\citep{Davidson1976}. For example, it has been adopted
by the World Chess Federation and the European Go Federation to rank players
and it is a standard approach to build multiclass classifiers based on the
output of binary classifiers~\citep{Hastie1998}. Various extensions have been
proposed to handle home advantage~\citep{Agresti1990}, draws~\citep{Rao1967},
multiple~\citep{Plackett1975,Luce1959} and team comparisons~\citep{Huang2006}.
In particular, the popular extension to multiple comparisons, named the
Plackett-Luce model \citep{Plackett1975,Luce1959}, defines a prior distribution
over permutations and has been used for ranking of multiple individuals and
for choice models~\citep{Luce1977}.\ The monographs of \cite{David1988} and \citet[Chap. 9]{Diaconis1988} provide detailed discussions on
the statistical foundations of these models.

For the basic Bradley-Terry model (\ref{eq:BTmodel}), it is possible to find
the maximum likelihood (ML) estimate of the skill ratings $\lambda$ using a
simple iterative procedure~\citep{Zermelo1929,Hunter2004}. \cite{Lange2000} established that this procedure is a specific case of the
general class of algorithms referred to as MM algorithms. Generally speaking,
MM algorithms use surrogate minimizing functions of the log-likelihood to
define an iterative procedure converging to a local maximum. EM\ algorithms
are thus just a special case of MM\ algorithms. An excellent survey of the
MM\ approach and its applications can be found in \citep{Lange2000}. \cite{Hunter2004} further derived MM algorithms for generalized
Bradley-Terry models and established sufficient conditions under which these
algorithms are guaranteed to converge towards the ML estimate.

Recently several authors have proposed to perform Bayesian inference for
(generalized) Bradley-Terry models
\citep{Adams2005,Gormley2009,Gorur2006,Guiver2009}. The resulting posterior
density is typically not tractable and needs to be approximated. An
Expectation-Propagation method is developed in \citep{Guiver2009}; this yields
an approximation of the posterior which can be computed quickly and might be
suitable for very large scale applications. However, it relies on a functional
approximation of the posterior and the convergence properties of this
algorithm are not well-understood. M-H algorithms have been proposed in
\citep{Adams2005,Gormley2009,Gorur2006}. \cite{Gormley2009} suggested a carefully
designed proposal distribution, though it can perform poorly in
some scenarios as demonstrated in section \ref{sec:experiments}.

Our contribution here is three-fold. First, we show that by introducing
suitable sets of latent variables, the MM algorithms proposed
by \cite{Hunter2004} for the basic Bradley-Terry model and its generalizations
to take into account home advantage, ties and multiple comparisons can be
reinterpreted as standard EM algorithms. We believe that this non-trivial
reinterpretation is potentially fruitful for statisticians who usually like
thinking in terms of latent variables. Note that the latent variables
introduced here differ from the ones introduced in the standard Thurstonian
interpretation~of the Bradley-Terry model \citep[Chap. 9]{Diaconis1988} and
lead to more efficient algorithms as discussed in section \ref{sec:bt}.
Second, using similar ideas, we propose original EM algorithms for some recent
generalizations of the Bradley-Terry model including group comparisons and
random graphs. Third, based on the sets of latent variables introduced to
derive these EM algorithms, we propose Gibbs samplers to perform Bayesian
inference in this important class of models. To the best of our knowledge, no
Gibbs sampler has ever been proposed in this context. These algorithms have
the great advantage of allowing us to bypass the design of proposal
distributions for M-H updates and we demonstrate experimentally that they
perform very well.

The rest of this paper is organized as follows. In section \ref{sec:bt}, we
consider the basic Bradley-Terry model (\ref{eq:BTmodel}). Based on the
introduction of a suitable set of latent variables, we present an EM
reinterpretation of the MM\ algorithm presented by \cite{Hunter2004} for
Maximum a Posteriori (MAP) parameter estimation and an original data
augmentation algorithm to sample from the posterior. In section
\ref{sec:genbt}, various standard extensions of the Bradley-Terry model
allowing for home advantage, ties and competition between teams are described.
EM\ algorithms and original Gibbs sampling schemes are proposed. The
Plackett-Luce model \citep{Plackett1975,Luce1959}, a very popular
generalization of the Bradley-Terry model for multiple comparisons, is
presented in section \ref{sec:plackettluce}. Algorithms applicable to further
extensions of the Bradley-Terry model to choice models, random graphs and
classification are presented in section \ref{sec:discussion}. In section
\ref{sec:experiments}, these algorithms are applied to the NASCAR 2002 dataset
and to chess competition data.

\section{Bradley-Terry model\label{sec:bt}}

Suppose we have observed a number of statistically independent pairwise
comparisons among $K$ individuals. We denote by $D$ the associated data. Let
also $w_{ij}$ denote the number of comparisons where $i$ beats $j$ and
$n_{ij}=w_{ij}+w_{ji}$ the total number of comparisons between $i$ and $j$.
Based on the Bradley-Terry model (\ref{eq:BTmodel}), the log-likelihood
function is given by%
\begin{align*}
\ell(\lambda) &  =\sum_{1\leq i\neq j\leq K}\text{ }w_{ij}\log\lambda
_{i}-w_{ij}\log(\lambda_{i}+\lambda_{j})\\
&  =\sum_{1\leq i\neq j\leq K}\text{ }w_{ij}\log\lambda_{i}-\sum_{1\leq
i<j\leq K}\text{ }n_{ij}\log(\lambda_{i}+\lambda_{j})
\end{align*}
where the notation $1\leq i\neq j\leq K$ is an abusive notation to denote the
set $\left\{  \left(  i,j\right)  \in\left\{  1,...,K\right\}  ^{2}\right .$ $\left .\text{ such
that }i\neq j\right\}  $ and $1\leq i<j\leq K$ stands for $\left\{  \left(
i,j\right)  \in\left\{  1,...,K\right\}  ^{2}\text{ such that }i<j\right\}  $.

We seek to introduce latent variables which are such that the resulting
complete log-likelihood admits a simple form. It is well-known that the
Bradley-Terry model enjoys the following Thurstonian
interpretation~\cite[Chap. 9]{Diaconis1988}: for each pair $1\leq i<j\leq K$
and for each associated pair comparison $k=1,\ldots,n_{ij}$, let $Y_{ki}%
\sim\mathcal{E}(\lambda_{i})$ and $Y_{kj}\sim\mathcal{E}(\lambda_{j})$ where
$\mathcal{E}\left(  \varsigma\right)  $ is the exponential distribution of
rate parameter $\varsigma$ then
\[
\Pr(Y_{ki}<Y_{kj})=\frac{\lambda_{i}}{\lambda_{i}+\lambda_{j}}.
\]
These latent variables can be interpreted as arrival times and the individual
with the lowest arrival time wins.\ These latent variables would allow us to
define EM and data augmentation algorithms. However, if we only introduce for
each pair $i,j$ the sum of the lowest arrival times
\[
Z_{ij}=\sum_{k=1}^{n_{ij}}\min(Y_{kj},Y_{ki})\sim\mathcal{G(}n_{ij}%
,\lambda_{i}+\lambda_{j})
\]
instead of $\left\{  Y_{ki},Y_{kj}\right\}  $ for $k=1,\ldots,n_{ij}$, then
the resulting complete log-likelihood remains simple. Here $\mathcal{G}\left(
\alpha,\beta\right)  $ denote the Gamma distribution of shape $\alpha$ and
inverse scale $\beta$. As the fraction of missing information is reduced, this
leads to faster rates of convergence for the resulting EM and data
augmentation algorithms \cite[Chap. 6]{Liu2001}. We will essentially proceed
similarly to introduce latent variables for generalized Bradley-Terry models.

To summarize, for $1\leq i<j\leq K$ such that $n_{ij}>0$, we introduce the
latent variables $Z=\left\{  Z_{ij}\right\}  $ which are such that
\begin{equation}
p\left(  \left.  z\right\vert D,\lambda\right)  =%
{\displaystyle\prod\limits_{1\leq i<j\leq K|n_{ij}>0}}
\mathcal{G}(z_{ij};n_{ij},\lambda_{i}+\lambda_{j})\label{eq:conditionallatent}%
\end{equation}
The resulting complete data log-likelihood is given by%
\begin{equation}
\ell(\lambda,z)=\sum_{1\leq i\neq j\leq K|w_{ij}>0}w_{ij}\log\lambda_{i}%
-\sum_{1\leq i<j\leq K|n_{ij}>0}(\lambda_{i}+\lambda_{j})z_{ij}+(n_{ij}-1)\log
z_{ij}-\log\Gamma(n_{ij})\label{eq:completelikelihood}%
\end{equation}
where $\Gamma$ is the Gamma function.

If we assign additionally a prior to $\lambda$ such that
\begin{equation}
p\left(  \lambda\right)  =%
{\displaystyle\prod\limits_{i=1}^{K}}
\mathcal{G}(\lambda_{i};a,b) \label{eq:prior}%
\end{equation}
as in \citep{Gormley2009,Guiver2009} then we can maximize the resulting
log-posterior using the EM\ algorithm which proceeds as follows at iteration
$t$:%
\begin{equation}
\lambda^{(t)}=\underset{\lambda}{\arg\max}\text{ }Q(\lambda,\lambda^{\left(
t-1\right)  }). \label{eq:maxEM}%
\end{equation}
We have
\begin{align}
Q(\lambda,\lambda^{\ast})  &  =\mathbb{E}_{\left.  Z\right\vert D,\lambda
^{\ast}}\left[  \ell(\lambda,Z)\right]  +\log\text{ }p\left(  \lambda\right)
\label{eq:QBT}\\
&  \equiv\sum_{i=1}^{K}\left(  a-1+w_{i}\right)  \log\lambda_{i}-b\lambda
_{i}-\sum_{1\leq i<j\leq K}(\lambda_{i}+\lambda_{j})\frac{n_{ij}}{\lambda
_{i}^{\ast}+\lambda_{j}^{\ast}}\nonumber
\end{align}
with \textquotedblleft$\equiv$\textquotedblright\ meaning \textquotedblleft
equal up to terms independent of the first argument of the $Q$
function\textquotedblright\ and $w_{i}=\sum_{j=1,j\neq i}^{K}w_{ij}$ is the
number of wins of element $i$. Using (\ref{eq:maxEM}), it follows that%
\begin{equation}
\lambda_{i}^{(t)}=\frac{a-1+w_{i}}{b+\sum_{j\neq i}\frac{n_{ij}}{\lambda
_{i}^{(t-1)}+\lambda_{j}^{(t-1)}}}. \label{eq:BTEMupdate}%
\end{equation}
For $a=1$ and $b=0$, the MAP and ML\ estimates coincide. In this case
(\ref{eq:QBT}) is exactly the minorizing function of the MM algorithm~proposed
in \cite[Eq. (10)]{Hunter2004} and thus the MM algorithm is given by
(\ref{eq:BTEMupdate}).

Based on the same latent variables, we present a simple data augmentation
algorithm for sampling from the posterior distribution $p\left(  \left.
\lambda,z\right\vert D\right)  \propto p\left(  \lambda\right)  \exp\left(
\ell(\lambda,z)\right)  $. By construction, we can update $Z$ conditional upon
$\lambda$ using (\ref{eq:conditionallatent}) and the conditional for $\lambda$
given $Z$ can be expressed easily so that the data augmentation sampler at
iteration $t$ proceeds as follows:

\begin{itemize}
\item For $1\leq i<j\leq K$ s.t. $n_{ij}>0$, sample
\[
Z_{ij}^{(t)}|D,\lambda^{(t-1)}\sim\mathcal{G}(n_{ij},\lambda_{i}%
^{(t-1)}+\lambda_{j}^{(t-1)}).
\]

\item For $i=1,\ldots,K,$ sample
\[
\lambda_{i}^{(t)}|D,Z^{(t)}\sim\mathcal{G(}a+w_{i},b+\sum_{i<j|n_{ij}>0}%
Z_{ij}^{(t)}+\sum_{i>j|n_{ij}>0}Z_{ji}^{(t)}).
\]
\medskip
\end{itemize}

\section{Generalized Bradley-Terry models\label{sec:genbt}}

\subsection{Home advantage}

Consider now that the pairwise comparisons are modeled using the Bradley-Terry
model with \textquotedblleft home-field advantage" \citep{Agresti1990} where%
\[
\Pr(i\text{ beats }j)=\left\{
\begin{array}
[c]{ll}%
\frac{\theta\lambda_{i}}{\theta\lambda_{i}+\lambda_{j}} & \text{if }i\text{ is
home,}\\
\frac{\lambda_{i}}{\lambda_{i}+\theta\lambda_{j}} & \text{if }j\text{ is
home.}%
\end{array}
\right.
\]
The parameter $\theta$, $\theta>0$, measures the strength of the home-field
advantage ($\theta>1$) or disadvantage ($\theta<1$). Let $a_{ij}$ be the
number of times that $i$ is at home and beats $j$ and $b_{ij}$ is the number
of times that $i$ is at home and loses to $j$.

The log-likelihood of the skill ratings $\lambda$ and $\theta$ is given by%
\[
\ell(\lambda,\theta)=c\log\theta+\sum_{i=1}^{K}w_{i}\log\lambda_{i}%
-\sum_{1\leq i\neq j\leq K}n_{ij}\log(\theta\lambda_{i}+\lambda_{j})
\]
where $n_{ij}=a_{ij}+b_{ij}$ is the number of times $i$ plays at home against
$j$, $c=\sum_{1\leq i\neq j\leq K}a_{ij}$ is the total number of home-field
wins and $w_{i}$ is the total number of wins of element $i$.

For $1\leq i\neq j\leq K$ such that $n_{ij}>0$, let us introduce the latent
variables $Z=\left\{  Z_{ij}\right\}  $ which are such that
\begin{equation}
p\left(  \left.  z\right\vert D,\lambda\right)  =%
{\displaystyle\prod\limits_{1\leq i\neq j\leq K|n_{ij}>0}}
\mathcal{G}(z_{ij};n_{ij},\theta\lambda_{i}+\lambda_{j}).
\end{equation}
The associated complete data log-likelihood is given by
\[
\ell(\lambda,\theta,z)=c\log\theta+\sum_{i=1}^{K}w_{i}\log\lambda_{i}%
-\sum_{1\leq i\neq j\leq K|n_{ij}>0}(\theta\lambda_{i}+\lambda_{j})z_{ij}%
+\log\Gamma\left(  n_{ij}\right)  .
\]

Using independent priors for $\lambda$ and $\theta$, i.e. $p\left(
\lambda,\theta\right)  =p\left(  \lambda\right)  p\left(  \theta\right)  $,
where $p\left(  \lambda\right)  $ is defined as (\ref{eq:prior}) and
\begin{equation}
\theta\sim\mathcal{G}(a_{\theta},b_{\theta}), \label{eq:priortheta}%
\end{equation}
then we can maximize the resulting posterior using the EM\ algorithm
\[
\left(  \lambda^{(t)},\theta^{\left(  t\right)  }\right)  =\text{
}\underset{\left(  \lambda,\theta\right)  }{\arg\max}\text{ }Q(\left(
\lambda,\theta\right)  ,\left(  \lambda^{\left(  t-1\right)  },\theta^{\left(
t-1\right)  }\right)  )
\]
where we have
\begin{align*}
Q(\left(  \lambda,\theta\right)  ,\left(  \lambda^{\ast},\theta^{\ast}\right)
)  &  =\mathbb{E}_{Z|D,\lambda^{\ast},\theta^{\ast}}\left[  \ell
(\lambda,\theta,Z)\right]  +\log\text{ }p\left(  \lambda,\theta\right) \\
&  \equiv\left(  a_{\theta}-1+c\right)  \log\theta-b_{\theta}\theta+\sum
_{i=1}^{K}\left(  a-1+w_{i}\right)  \log\lambda_{i}-b\sum_{i=1}^{K}\lambda
_{i}-\sum_{1\leq i\neq j\leq K}n_{ij}\frac{\theta\lambda_{i}+\lambda_{j}%
}{\theta^{\ast}\lambda_{i}^{\ast}+\lambda_{j}^{\ast}}.
\end{align*}
We obtain
\begin{align*}
\lambda_{i}^{(t)}  &  =\frac{a-1+w_{i}}{b+\sum_{1\leq i\neq j\leq K}\text{
}\left\{  \frac{\theta^{(t-1)}n_{ij}}{\theta^{(t-1)}\lambda_{i}^{(t-1)}%
+\lambda_{j}^{(t-1)}}+\frac{n_{ji}}{\theta^{(t-1)}\lambda_{j}^{(t-1)}%
+\lambda_{i}^{(t-1)}}\right\}  }\text{ for }i=1,\ldots,K,\\
\theta^{(t)}  &  =\frac{a_{\theta}-1+c}{b_{\theta}+\sum_{1\leq i\neq j\leq
K}\frac{n_{ij}\text{ }\lambda_{i}^{(t)}}{\theta^{(t-1)}\lambda_{i}%
^{(t-1)}+\lambda_{j}^{(t-1)}}}.
\end{align*}
For $a=a_{\theta}=1$ and $b=b_{\theta}=0$, i.e. if we use flat priors, this EM
algorithm is similar to the MM algorithm~proposed in \cite[pp. 389]%
{Hunter2004}.

Using the same latent variables, we can sample from the posterior distribution
of $\left(  \lambda,\theta,Z\right)  $ using the Gibbs sampler which updates
iteratively $Z,$ $\lambda$ and $\theta$ as follows at iteration $t$:

\begin{itemize}
\item For $1\leq i\neq j\leq K$ s.t. $n_{ij}>0$, sample
\[
Z_{ij}^{(t)}|D,\lambda^{(t-1)},\theta^{(t-1)}\sim\mathcal{G}\left(
n_{ij},\theta^{(t-1)}\lambda_{i}^{(t-1)}+\lambda_{j}^{(t-1)}\right)  .
\]

\item For $i=1,\ldots,K,$ sample
\[
\lambda_{i}^{(t)}|D,\theta^{(t-1)},Z^{(t)}\sim\mathcal{G}\left(
a+w_{i},b+\theta^{\left(  t-1\right)  }\sum_{j\neq i|n_{ij}>0}Z_{ij}%
^{(t)}+\sum_{j\neq i|n_{ij}>0}Z_{ji}^{\left(  t\right)  }\right)  .
\]

\item Sample
\[
\theta^{\left(  t\right)  }|D,\lambda^{(t)},Z^{(t)}\sim\mathcal{G}\left(
a_{\theta}+c,b_{\theta}+\sum_{i=1}^{K}\lambda_{i}^{\left(  t\right)  }%
\sum_{j\neq i|n_{ij}>0}Z_{ij}^{\left(  t\right)  }\right)  .
\]

\end{itemize}

\subsection{Model with ties\label{sec:ties}}

If we now want to allow for ties in pairwise comparisons, we can use the
following model proposed by \cite{Rao1967}
\begin{align*}
\Pr(i\text{ beats }j)  &  =\frac{\lambda_{i}}{\lambda_{i}+\theta\lambda_{j}%
},\\
\Pr(i\text{ ties }j)  &  =\frac{(\theta^{2}-1)\lambda_{i}\lambda_{j}}%
{(\lambda_{i}+\theta\lambda_{j})(\theta\lambda_{i}+\lambda_{j})}%
\end{align*}
where $\theta>1$. The log-likelihood function for $\left(  \lambda
,\theta\right)  $ is given by%
\begin{align*}
\ell(\lambda,\theta)  &  =\sum_{1\leq i\neq j\leq K}w_{ij}\log\frac
{\lambda_{i}}{\lambda_{i}+\theta\lambda_{j}}+\frac{t_{ij}}{2}\log\frac
{(\theta^{2}-1)\lambda_{i}\lambda_{j}}{(\theta\lambda_{i}+\lambda_{j}%
)(\lambda_{i}+\theta\lambda_{j})}\\
&  =\sum_{1\leq i\neq j\leq K}s_{ij}\log\frac{\lambda_{i}}{\lambda_{i}%
+\theta\lambda_{j}}+\frac{t_{ij}}{2}\log(\theta^{2}-1)
\end{align*}
where $t_{ij}=t_{ji}$ is the number of ties between $i$ and $j$ and
$s_{ij}=w_{ij}+t_{ij}$.

For $1\leq i\neq j\leq K$ such that $s_{ij}>0$, let us introduce the latent
variables $Z=\left\{  Z_{ij}\right\}  $ which are such that
\[
p\left(  \left.  z\right\vert D,\lambda\right)  =%
{\displaystyle\prod\limits_{1\leq i\neq j\leq K|s_{ij}>0}}
\mathcal{G}(z_{ij};s_{ij},\lambda_{i}+\theta\lambda_{j})
\]
which yields the following complete log-likelihood%
\[
\ell(\lambda,\theta,z)=T\log(\theta^{2}-1)+\sum_{1\leq i\neq j\leq K|s_{ij}%
>0}s_{ij}\log\lambda_{i}-(\lambda_{i}+\theta\lambda_{j})z_{ij}+(s_{ij}-1)\log
z_{ij}-\log\Gamma(s_{ij})
\]
\bigskip where $T=\frac{1}{2}\sum_{1\leq i\neq j\leq K}t_{ij}$ is the total
number of ties. If we adopt for $\theta$ a flat improper prior on $\left[
1,\infty\right)  $ and we select $p\left(  \lambda\right)  $ as
(\ref{eq:prior}) then we obtain%
\begin{align*}
Q(\left(  \lambda,\theta\right)  ,\left(  \lambda^{\ast},\theta^{\ast}\right)
)  &  =\mathbb{E}_{Z|D,\lambda^{\ast},\theta^{\ast}}\left[  \ell
(\lambda,\theta,Z)\right]  +\log\text{ }p\left(  \lambda,\theta\right) \\
&  \equiv T\log(\theta^{2}-1)+\sum_{1\leq i\neq j\leq K}s_{ij}\left(
\log\lambda_{i}-\frac{\lambda_{i}+\theta\lambda_{j}}{\lambda_{i}^{\ast}%
+\theta^{\ast}\lambda_{j}^{\ast}}\right)  +\sum_{i=1}^{K}\left(  a-1\right)
\log\lambda_{i}-b\lambda_{i}%
\end{align*}
and we recover once again the minorizing function in \cite[pp. 389-390]%
{Hunter2004} for $a=1$ and $b=0$. This can be maximized using the following procedure

\begin{itemize}
\item For $i=1,\ldots,K,$ set
\[
\lambda_{i}^{(t)}=\left(  a-1+\sum_{j\neq i}s_{ij}\right)  \left[
b+\sum_{j\neq i}\frac{s_{ij}}{\lambda_{i}^{(t-1)}+\theta^{(t-1)}\lambda
_{j}^{(t-1)}}+\frac{\theta^{(t-1)}s_{ji}}{\theta^{(t-1)}\lambda_{i}%
^{(t-1)}+\lambda_{j}^{(t-1)}}\right]  ^{-1}\text{.}%
\]

\item Set
\[
\theta^{(t)}=\frac{1}{2c^{(t)}}+\sqrt{1+\frac{1}{4c^{(t)\text{ }2}}}%
\]
where
\[
c^{(t)}=\sum_{1\leq i\neq j\leq K}\frac{s_{ij}\lambda_{j}^{(t)}}{\lambda
_{i}^{(t-1)}+\theta^{(t-1)}\lambda_{j}^{(t-1)}}\text{.}%
\]

\end{itemize}

Using the same latent variables, we can sample from the posterior distribution
of $\left(  \lambda,\theta,Z\right)  $ using the following Gibbs sampler which
updates iteratively $Z,$ $\lambda$ and $\theta$ as follows at iteration $t$:

\begin{itemize}
\item For $1\leq i\neq j\leq K$ s.t. $s_{ij}>0$, sample
\[
Z_{ij}^{\left(  t\right)  }|D,\lambda^{(t-1)},\theta^{(t-1)}\sim
\mathcal{G}\left(  s_{ij},\lambda_{i}^{\left(  t-1\right)  }+\theta
^{(t-1)}\lambda_{j}^{\left(  t-1\right)  }\right)  .
\]

\item For $i=1,\ldots,K$, sample%
\[
\lambda_{i}^{(t)}|D,\theta^{(t-1)},Z^{(t)}\sim\mathcal{G}\left(  a+\sum_{j\neq
i}s_{ij},b+\sum_{j\neq i|s_{ij}>0}Z_{ij}^{\left(  t\right)  }+\theta
^{(t-1)}\sum_{j\neq i|s_{ij}>0}Z_{ji}^{\left(  t\right)  }\right)  .
\]

\item Sample
\[
\theta^{\left(  t\right)  }|D,\lambda^{(t)},Z^{(t)}\sim p(\theta
|D,\lambda^{(t)},Z^{\left(  t\right)  })
\]
where%
\begin{equation}
p(\theta|D,Z,\lambda)\propto(\theta^{2}-1)^{T}\exp\left(  -\sum_{1\leq i\neq
j\leq K|s_{ij}>0}Z_{ij}\text{ \ }\theta\right)  1_{\theta>1}.
\label{eq:conditionalnonstandardtheta}%
\end{equation}

\end{itemize}

It is possible to sample from (\ref{eq:conditionalnonstandardtheta}) exactly.
By performing a change of variable $\overline{\theta}=\theta-1$, we obtain
\[
p(\overline{\theta}|D,Z,\lambda)\propto(\overline{\theta}^{2}+2\overline
{\theta})^{T}\exp\left(  -\sum_{1\leq i\neq j\leq K|s_{ij}>0}Z_{ij}\text{
\ }\overline{\theta}\right)
\]
which is a mixture of Gamma distributions.

\subsection{Group comparisons}

Consider now that we have $n$ pairwise comparisons betweem teams. For each
comparison $i=1,\ldots,n$, let \linebreak$T_{i}^{+}\subset\{1,\ldots,K\}$ be
the winning team and $T_{i}^{-}\subset\{1,\ldots,K\}$ the losing team where
$T_{i}^{+}\cap T_{i}^{-}=\varnothing$ and $T_{i}=T_{i}^{+}\cup T_{i}^{-}$.
Recently \cite{Huang2006} have proposed the following model%
\begin{equation}
\Pr(T_{i}^{+}\text{ beats }T_{i}^{-})=\frac{\sum_{j\in T_{i}^{+}}\lambda_{j}%
}{\sum_{j\in T_{i}}\lambda_{j}}\text{.}%
\end{equation}
The log-likelihood function for $\lambda$ is thus given by
\[
\ell(\lambda)=\sum_{i=1}^{n}\log\left(  \sum_{j\in T_{i}^{+}}\lambda
_{j}\right)  -\log\left(  \sum_{j\in T_{i}}\lambda_{j}\right)  \text{.}%
\]

For $i=1,...,n$ we introduce the latent variables $Z=\left\{  Z_{i}\right\}  $
and $C=\left\{  C_{i}\right\}  $ such that
\[
p\left(  \left.  z,c\right\vert D,\lambda\right)  =p\left(  \left.
z\right\vert D,\lambda\right)  P\left(  \left.  c\right\vert D,\lambda\right)
\]
with%
\begin{align*}
p\left(  \left.  z\right\vert D,\lambda\right)   &  =%
{\displaystyle\prod\limits_{i=1}^{n}}
\mathcal{E}\left(  z_{i};\sum_{j\in T_{i}}\lambda_{j}\right)  ,\\
\Pr\left(  \left.  c\right\vert D,\lambda\right)   &  =%
{\displaystyle\prod\limits_{i=1}^{n}}
\frac{\lambda_{c_{i}}}{\sum_{j\in T_{i}^{+}}\lambda_{j}}\text{ with }c_{i}\in
T_{i}^{+}%
\end{align*}
where $\mathcal{E}\left(  x;\alpha\right)  $ is the exponential density of
argument $x$ and parameter $\alpha$. It follows that the complete
log-likelihood is given by%
\[
\ell(\lambda,z,c)=\sum_{i=1}^{n}\log\lambda_{c_{i}}-\left(  \sum_{j\in T_{i}%
}\lambda_{j}\right)  z_{i}.
\]

The $Q$ function associated to the EM algorithm is given by
\begin{align*}
Q(\lambda,\lambda^{\ast})  &  =\mathbb{E}_{Z,C|D,\lambda^{\ast}}\left[
\ell(\lambda,Z,C)\right]  +\log\text{ }p\left(  \lambda\right) \\
&  \equiv\sum_{i=1}^{n}\sum_{j\in T_{i}^{+}}\log\lambda_{j}\frac{\lambda
_{j}^{\ast}}{\sum_{k\in T_{i}^{+}}\lambda_{k}^{\ast}}-\frac{\sum_{j\in T_{i}%
}\lambda_{j}}{\sum_{j\in T_{i}}\lambda_{j}^{\ast}}+\sum_{k=1}^{K}\left(
a-1\right)  \log\lambda_{k}-b\lambda_{k}\\
&  \equiv\sum_{k=1}^{K}\left(  a-1+\lambda_{k}^{\ast}\sum_{i=1}^{n}%
\frac{\alpha_{ik}}{\sum_{j\in T_{i}^{+}}\lambda_{j}^{\ast}}\right)
\log\lambda_{k}-\lambda_{k}\left(  b+\sum_{i=1}^{n}\frac{\gamma_{ik}}%
{\sum_{j\in T_{i}}\lambda_{j}^{\ast}}\right)
\end{align*}
where $\alpha_{ik}=1$ if $k\in T_{i}^{+}$ and $0$ otherwise and $\gamma
_{ik}=1$ if $k\in T_{i}$ and $0$ otherwise. It follows that the EM update is
given by%
\[
\lambda_{k}^{\left(  t\right)  }=\frac{a-1+\lambda_{k}^{\ast}\sum_{i=1}%
^{n}\frac{\alpha_{ik}}{\sum_{j\in T_{i}^{+}}\lambda_{j}^{\left(  t-1\right)
}}}{b+\sum_{i=1}^{n}\frac{\gamma_{ik}}{\sum_{j\in T_{i}}\lambda_{j}^{\left(
t-1\right)  }}}.
\]

Using the same latent variables, we obtain a data augmentation sampler to
sample from $p\left(  \left.  \lambda,z,c\right\vert D\right)  $ by
iteratively sampling $\left(  Z,C\right)  $ and $\lambda.$ This proceeds as
follows at iteration $t$:

\begin{itemize}
\item For $i=1,...,n$, sample
\[%
\begin{tabular}
[c]{l}%
$Z_{i}^{\left(  t\right)  }|D,\lambda^{(t-1)}\sim\mathcal{E}\left(  \sum_{j\in
T_{i}}\lambda_{j}^{\left(  t-1\right)  }\right)  ,$\\
$\Pr\left(  C_{i}^{\left(  t\right)  }=k|D,\lambda^{(t-1)}\right)
=\frac{\lambda_{k}^{\left(  t-1\right)  }}{\sum_{j\in T_{i}^{+}}\lambda
_{j}^{\left(  t-1\right)  }},$ $k\in T_{i}^{+}.$%
\end{tabular}
\ \
\]

\item For $k=1,\ldots,K$, sample%
\[
\lambda_{k}^{(t)}|D,Z^{(t)},C^{(t)}\sim\mathcal{G}\left(  a+\sum_{i=1}%
^{n}\delta_{k,C_{i}^{\left(  t\right)  }},b+\sum_{i=1}^{n}\gamma_{ik}%
Z_{i}^{\left(  t\right)  }\right)
\]
where $\delta_{u,v}=1$ if $u=v$ and $0$ otherwise.
\end{itemize}

\section{Multiple comparisons\label{sec:plackettluce}}

We now consider the popular Plackett-Luce model \citep{Luce1959,Plackett1975}
which is an extension of the Bradley-Terry model to comparisons involving more
than two elements. Assume that $p_{i}\leq K$ individuals are ranked for
comparison $i$ where $i=1,...,n$. We write $\rho_{i}=(\rho_{i1},\ldots
,\rho_{ip_{i}})$ where $\rho_{i1}$ is the first individual, $\rho_{i2}$, the
second, etc. The Plackett-Luce model assumes%
\begin{equation}
\Pr(\rho_{i}|\lambda)=\prod_{j=1}^{p_{i}}\frac{\lambda_{\rho_{ij}}}{\sum
_{k=j}^{p_{i}}\lambda_{\rho_{ik}}}=\prod_{j=1}^{p_{i}-1}\frac{\lambda
_{\rho_{ij}}}{\sum_{k=j}^{p_{i}}\lambda_{\rho_{ik}}}. \label{eq:plackettluce}%
\end{equation}

For $i=1,\ldots,n$ and $j=1,\ldots,p_{i}-1$, we introduce the following latent
variables $Z=\left\{  Z_{ij}\right\}  $
\[
p\left(  \left.  z\right\vert D,\lambda\right)  =\prod_{i=1}^{n}\prod
_{j=1}^{p_{i}-1}\mathcal{E(}z_{ij};\sum_{k=j}^{p_{i}}\lambda_{\rho_{ik}})
\]
which leads to the complete log-likelihood
\[
\ell(\lambda,z)=\sum_{i=1}^{n}\sum_{j=1}^{p_{i}-1}\text{ }\log\lambda
_{\rho_{ij}}-\left(  \sum_{k=j}^{p_{i}}\lambda_{\rho_{ik}}\right)  z_{ij}.
\]

The $Q$ function associated to the EM algorithm is given by
\begin{align*}
Q(\lambda,\lambda^{\ast})  &  =\mathbb{E}_{Z|D,\lambda^{\ast}}\left[
\ell(\lambda,Z)\right]  +\log\text{ }p\left(  \lambda\right) \\
&  \equiv\sum_{i=1}^{n}\sum_{j=1}^{p_{i}-1}\log\lambda_{\rho_{ij}}-\frac
{\sum_{k=j}^{p_{i}}\lambda_{\rho_{ik}}}{\sum_{k=j}^{p_{i}}\lambda_{\rho_{ik}%
}^{\ast}}+\sum_{k=1}^{K}\text{ }\left(  a-1\right)  \log\lambda_{k}%
-b\lambda_{k}%
\end{align*}
which is once again equivalent to the majorizing function in \cite[pp.
398]{Hunter2004} for $a=1,$ $b=0$. It follows that the EM\ algorithm is given
at iteration $t$ by
\[
\lambda_{k}^{(t)}=(a-1+w_{k})\left[  b+\sum_{i=1}^{n}\left(  \sum_{j=1}%
^{p_{i}-1}\frac{\delta_{ijk}}{\sum_{k=j}^{p_i}\lambda_{\rho_{ik}}^{(t-1)}%
}\right)  \right]  ^{-1}%
\]
where $w_{k}$ is the number of rankings where the $k^{\text{th}}$ individual
is not in the last ranking position and $\delta_{ijk}$ is defined as%
\[
\delta_{ijk}=\left\{
\begin{array}
[c]{ll}%
1 & \text{if }k\in\{\rho_{ij},\ldots,\rho_{ip_{j}}\}\\
0 & \text{otherwise}%
\end{array}
\right.
\]
i.e. $\delta_{ijk}$ is the indicator of the event that individual $k$ receives
a rank no better than $j$ in the $i^{\text{th}}$ ranking.

To sample from $p\left(  \left.  \lambda,z\right\vert D\right)  $, we can use
the following data augmentation sampler. At iteration $t$, this proceeds as follows:

\begin{itemize}
\item For $i=1,...,n$, for $j=1,\ldots,p_{i}-1$, sample%
\[
Z_{ij}^{\left(  t\right)  }|D,\lambda^{\left(  t-1\right)  }\sim
\mathcal{E}(\sum_{k=j}^{p_{i}}\lambda_{\rho_{ik}}^{\left(  t-1\right)  }).
\]

\item For $k=1,\ldots,K$, sample
\[
\lambda_{k}^{\left(  t\right)  }|D,Z^{\left(  t\right)  }\sim\mathcal{G}%
(a+w_{k},b+\sum_{i=1}^{n}\sum_{j=1}^{p_{i}-1}\delta_{ijk}Z_{ij}^{\left(
t\right)  }).
\]

\end{itemize}

Using exactly the same augmentation, EM and Gibbs samplers can be defined for
further extensions of these models such as mixtures of Plackett-Luce
models~\citep{Gormley2008a}.

\section{Discussion\label{sec:discussion}}

\subsection{Identifiability}

Consider the basic Bradley-Terry model and its extensions to group comparisons
and multiple comparisons. Let us define
\[
\Lambda=\sum_{i=1}^{K}\lambda_{i}\text{ \ and }\pi_{i}=\frac{\lambda_{i}%
}{\Lambda}%
\]
and write $\pi:=\left\{  \pi_{i}\right\}  _{i=1}^{K}$.\ The likelihood is
invariant to a rescaling of the vector $\lambda$ so the parameter $\Lambda$ is
not likelihood-identifiable and
\[
p(\pi,\Lambda|D)=p(\left.  \pi\right\vert D)p(\Lambda).
\]
From (\ref{eq:prior}), it follows that $\pi\sim\mathcal{D}(a,\ldots,a)$ where
$\mathcal{D}$ is the Dirichlet distribution and $\Lambda\sim\mathcal{G}%
(Ka,b)$, hence
\[
\Lambda^{MAP}=\frac{aK-1}{b}.
\]

To improve the mixing of the MCMC\ algorithms in this context, an additional
sampling step can be added where we normalize the current parameter estimate
$\lambda^{\left(  t\right)  }$ and then rescale them randomly using a prior
draw for $\Lambda$.

\begin{itemize}
\item For $i=1,\ldots,K$, set $\lambda_{i}^{\ast(t)}=\frac{\lambda_{i}^{(t)}%
}{\sum_{j=1}^{K}\lambda_{j}^{(t)}}\Lambda^{(t)}$ where $\Lambda^{(t)}%
\sim\mathcal{G}(Ka,b)$.
\end{itemize}

This step can drastically improve the mixing of the Markov chain. However, if
we are only interested in the normalized values $\pi$ of $\lambda$ then this
additional step is useless.

As an alternative, it is also possible to consider an EM\ algorithm for the
basic Bradley-Terry model which does not require the introduction of a scale
parameter. Assume $\pi\sim\mathcal{D}(a,\ldots,a)$ and let us introduce latent
variables $M_{ij}$, $C_{ij}=(C_{ij1},\ldots,C_{ijM_{ij}})$ for $1\leq i\neq
j\leq K$ such that $n_{ij}>0$
\begin{align*}
M_{ij} &  \sim NB(n_{ij},\pi_{i}+\pi_{j}),\\
\Pr\left(  C_{ijk}=l\right)   &  =\frac{\pi_{l}}{\sum_{n\neq i,j}\pi_{n}%
}\text{ for }k=1,\ldots M_{ij}\text{ and }l\neq i,j
\end{align*}
where $NB(r,p)$ is the negative binomial distribution.\ The complete
log-likelihood is given by
\[
\ell(\pi,m,c)=\sum_{i=1}^{K}\sum_{j=1,j\neq i}^{K}w_{ij}\log\pi_{i}+\sum
_{i=1}^{K}\sum_{j=1,j\neq i}^{K}\sum_{k\neq i,j}\left[  \log\binom
{n_{ij}+m_{ij}-1}{n_{ij}-1}+r_{ijk}\log\pi_{k}\right]
\]
where $r_{ijk}$ is the number of $c_{ijl},l=1,\ldots,m_{ij}$ that take value
$k$. Omitting the terms independent of $\pi$, the $Q$ function is given by%
\begin{align*}
Q(\pi,\pi^{\ast}) &  =\mathbb{E}_{M|D,\pi^{\ast}}\left[  \mathbb{E}%
_{C|D,M,\pi^{\ast}}\left[  \ell(\pi,M,C)\right]  \right]  +\log\text{
}p\left(  \pi\right)  \\
&  \equiv\mathbb{E}_{M|D,\pi^{\ast}}\left[  \sum_{i=1}^{K}\sum_{j=1,j\neq
i}^{K}w_{ij}\log\pi_{i}+\sum_{i=1}^{K}\sum_{j=1,j\neq i}^{K}\sum_{k\neq
i,j}\log\binom{n_{ij}+M_{ij}-1}{n_{ij}-1}+M_{ij}\frac{\pi_{k}^{\ast}}%
{\sum_{l\neq i,j}\pi_{l}^{\ast}}\log\pi_{k}\right]  \\
&+\log\text{ }p\left(
\pi\right)  \\
&  \equiv\sum_{i=1}^{K}\sum_{j=1,j\neq i}^{K}w_{ij}\log\pi_{i}+\sum_{i=1}%
^{K}\sum_{j=1,j\neq i}^{K}\sum_{k\neq i,j}n_{ij}\frac{\pi_{k}^{\ast}}{\pi
_{i}^{\ast}+\pi_{j}^{\ast}}\log\pi_{k}+\left(  a-1\right)  \sum_{i=1}^{K}%
\log\pi_{k}+C\\
&  \equiv\sum_{k=1}^{K}(w_{k}+\pi_{k}^{\ast}\sum_{i=1,i\neq k}^{K}%
\sum_{j=1,j\neq i,k}^{K}\frac{n_{ij}}{\pi_{i}^{\ast}+\pi_{j}^{\ast}}\log
\pi_{k})+\left(  a-1\right)  \sum_{i=1}^{K}\log\pi_{k}+C
\end{align*}
where $C$ is a term independent of $\pi$.\ It follows that the EM\ update is
given by%
\[
\pi_{k}^{(t)}\propto a-1+w_{k}+\pi_{k}^{(t-1)}\sum_{i=1,i\neq k}^{K}%
\sum_{j=1,j\neq i,k}^{K}\frac{n_{ij}}{\pi_{i}^{(t-1)}+\pi_{j}^{(t-1)}}%
\]
with $\sum_{k=1}^{K}\pi_{k}^{\left(  t\right)  }=1.$ Although the above
EM\ algorithm does not rely on unidentifiable scaling parameters, it suffers
from a slow convergence rate. When $\pi_{k}$ takes small values, $\sum_{i\neq
k}\sum_{j\neq k}\frac{n_{ij}}{\pi_{i}+\pi_{j}}$ is large and it slows down the
convergence of the EM\ algorithm. The same augmentation can be used to define
a Gibbs sampler, but the same slow mixing issues arise for the Markov chain.

\subsection{Hyperparameter estimation}

The prior (\ref{eq:prior}) is specified by the hyperparameters $a$ and $b$.
However, the inverse scale parameter $b$ is not likelihood identifiable so
there is no point assigning a prior to it. However it might be interesting to
set a prior $p(a)$ on $a$ and estimate it from the data. Given $\lambda$, we
have
\[
p\left(  \left.  a\right\vert \lambda\right)  \propto p\left(  a\right)
\underset{l_{1}\left(  a\right)  }{\underbrace{\left(  b^{K}\prod_{i=1}%
^{K}\lambda_{i}\right)  ^{a}\text{\ }}}\text{ }\underset{l_{2}%
(a)}{\underbrace{\Gamma^{-K}(a)}}.
\]
It is possible to sample from this density using auxiliary variables
$U_{1},U_{2}$ defined on $\left(  0,\infty\right)  $ as described in
\citep{Damien1999}. We introduce
\[
p\left(  \left.  a,u_{1},u_{2}\right\vert \lambda\right)  \propto p\left(
a\right)  \mathbb{I}\left\{  u_{1}<l_{1}\left(  a\right)  \right\}
\mathbb{I}\left\{  u_{2}<l_{2}\left(  a\right)  \right\}  .
\]
A Gibbs sampler can now be implemented to sample from $p\left(  \left.
a,u_{1},u_{2}\right\vert \lambda\right)  $. We can directly sample from the
full conditionals of $U_{1}$ and $U_{2}$
\[
\left.  U_{1}\right\vert \lambda\sim\mathcal{U}\left(  0,l_{1}\left(
a\right)  \right)  ,\text{ \ }\left.  U_{2}\right\vert \lambda\sim
\mathcal{U}\left(  0,l_{2}\left(  a\right)  \right)
\]
where $\mathcal{U}\left(  \alpha,\beta\right)  $ is the uniform distribution
on $\left(  \alpha,\beta\right)  $. The full conditional of $a$ given
$u_{1},u_{2}$ is given by
\[
p\left(  \left.  a\right\vert \lambda,u_{1},u_{2}\right)  \propto p\left(
a\right)  \mathbb{I}_{A_{1}\cap A_{2}}\left(  a\right)
\]
where
\[
A_{i}=\left\{  a:l_{i}\left(  a\right)  >u_{i}\right\}  .
\]
Alternatively we can update $a$ using a M-H random walk on $\log(a)$. We can
propose $a^{\star}=\exp(\sigma_{a}^{2}z)a$ where $z\sim\mathcal{N}(0,1)$ and
$a^{\star}$ is accepted with probability
\[
\min\left\{  1,\frac{p(a^{\star})}{p(a)}\left(  \frac{\Gamma(a)}%
{\Gamma(a^{\star})}\right)  ^{K}\left(  b^{K}\prod_{i=1}^{K}\lambda
_{i}\right)  ^{a^{\star}-a}\right\}  .
\]

\subsection{Further extensions}

\subsubsection{Random graphs with a given degree sequence}

A model closely related to Bradley-Terry has been proposed for undirected
random graphs with $K$ vertices~\citep{Holland1981,Chatterjee2010,Park2004}. In
this model, the degree sequence $(d_{1},\ldots,d_{K})$ of a given graph, where
$d_{i}$ is the degree of node $i$, is supposed to capture all the information
in the graph. It can be formalized by saying that the degree sequence is a
sufficient statistic for a probability distribution on
graphs~\citep{Chatterjee2010}.

In this model an edge is inserted between vertices $i$ and $j$ for $1\leq
i<j\leq K$ with probability%
\[
\frac{\lambda_{i}\lambda_{j}}{1+\lambda_{i}\lambda_{j}}%
\]
where $\lambda_{k}>0$ for $k\in\{1,\ldots,K\}$. Let $r_{ij}=1$ if there is an
edge between $i$ and $j$ and $0$ otherwise. Given the observations $D=\left\{
r_{ij}\right\}  _{1\leq i<j\leq K}$, the log-likelihood function for $\lambda$
is given by
\[
\ell(\lambda)=\sum_{1\leq i<j\leq K}r_{ij}\log\left(  \lambda_{i}\lambda
_{j}\right)  -\log\left(  1+\lambda_{i}\lambda_{j}\right)  .
\]

We introduce the following latent variables $Z=\left\{  Z_{ij}\right\}
_{1\leq i<j\leq K}$ such that%
\[
p\left(  \left.  z\right\vert D,\lambda\right)  =%
{\displaystyle\prod\limits_{1\leq i<j\leq K}}
\mathcal{E}(z_{ij};\lambda_{i}+\frac{1}{\lambda_{j}})\text{.}%
\]
The complete log-likelihood is given by%
\[
\ell(\lambda,z)=\sum_{1\leq i<j\leq K}r_{ij}\log\lambda_{i}-(1-r_{ij}%
)\log\lambda_{j}-(\lambda_{i}+\frac{1}{\lambda_{j}})z_{ij}%
\]

The $Q$ function associated to the EM algorithm is given by%
\begin{align*}
Q(\lambda,\lambda^{\ast})  &  =\mathbb{E}_{Z|D,\lambda^{\ast}}\left[
\ell(\lambda,Z)\right]  +\log\text{ }p\left(  \lambda\right) \\
&  \equiv\sum_{i=1}^{K}\log\lambda_{i}\left[  \left(  a-1\right)  +\sum
_{j>i}r_{ij}-\sum_{j<i}(1-r_{ij})\right]  -\lambda_{i}\left(  b+\sum
_{j>i}\frac{1}{\lambda_{i}^{\ast}+\frac{1}{\lambda_{j}^{\ast}}}\right)
-\frac{1}{\lambda_{i}}\sum_{j<i}\frac{1}{\lambda_{j}^{\ast}+\frac{1}%
{\lambda_{i}^{\ast}}}.
\end{align*}
Solving $\partial Q(\lambda,\lambda^{\ast})/\partial\lambda_{i}=0$ requires
solving a quadratic equation. For sake of brevity, we do not present these
details here.

Once again, we can define a data augmentation sampler to sample from $p\left(
\left.  \lambda,z\right\vert D\right)  $ by iteratively sampling $Z$ and
$\lambda.$ This proceeds as follows at iteration $t$:

\begin{itemize}
\item For $1\leq i<j\leq K$, sample%
\[
\left.  Z_{ij}^{\left(  t\right)  }\right\vert D,\lambda^{(t-1)}%
\sim\mathcal{E(}\lambda_{i}^{\left(  t-1\right)  }+\frac{1}{\lambda
_{j}^{\left(  t-1\right)  }}).
\]

\item For $i=1,...,K$, sample%
\[
\lambda_{i}^{\left(  t\right)  }|D,Z^{\left(  t\right)  }\sim\mathcal{GIG}%
\left(  2(\sum_{j>i}Z_{ij}^{\left(  t\right)  }+b),2\sum_{j<i}Z_{ij}^{\left(
t\right)  },a+\sum_{j>i}r_{ij}-\sum_{j<i}(1-r_{ij})\right)  .
\]

\end{itemize}

Here $\mathcal{GIG}\left(  \alpha,\beta,\gamma\right)  $ denotes the
generalized inverse Gaussian distribution (see e.g.
\citep{Barndorff-Nielsen2001}) whose density for an argument $x$ is
proportional to
\[
x^{\gamma-1}\exp\left\{  -\left(  \alpha x+\beta/x\right)  /2\right\}  .
\]
Algorithms to sample exactly from this distribution are available.

\subsubsection{Choice models}

Other extensions of the Bradley-Terry model are the choice models introduced
by \cite{Restle1961} and \cite{Tversky1972,Tversky1972a} in
psychology; see also~\citep{Wickelmaier2004,Gorur2006}. In these models, we are
given a set of $n$ elements. To each element $i$ is associated a set of $K$
features represented by a binary vector $f_{i}\in\{0,1\}^{K}$. The probability
that element $i$ is chosen over element $j$ is given by%
\[
\pi_{ij}=\frac{\sum_{k=1}^{K}\lambda_{k}f_{ik}(1-f_{jk})}{\sum_{k=1}%
^{K}\lambda_{k}f_{ik}(1-f_{jk})+\sum_{k=1}^{K}\lambda_{k}f_{jk}(1-f_{ik})}%
\]
where $\lambda_{k}>0$ is a weight representing the importance of feature $k$.
The term $\sum_{k=1}^{K}\lambda_{k}f_{ik}(1-f_{jk})$ corresponds to the sum of
the weights of features possessed by object $i$ but not object $j$. EM and
Gibbs algorithms can be derived by following the same construction as with
group comparisons.\bigskip

\subsubsection{Classification model}

Let consider the following original model for categorical data analysis%
\begin{align}
\Pr(Y_{i}=k) &  =\frac{\sum_{j=1}^{p}\exp(X_{ij})\lambda_{kj}}{\sum_{l=1}%
^{K}\sum_{j=1}^{p}\exp(X_{ij})\lambda_{lj}}\nonumber\\
&  =\frac{\exp(X_{i})^{T}\lambda_{k}}{\sum_{l=1}^{K}\exp(X_{i})^{T}\lambda
_{l}}\label{eq:classif}%
\end{align}
where $X_{i}\in\mathbb{R}^{p}$ is a vector of covariates and $\lambda_{k}%
\in\mathbb{R}_{+}^{p}$ for $k=1,\ldots,K$.\ This model could be used as an
alternative to the multinomial logit model~\citep{Agresti1990}. By introducing
latent variables $Z_{i}\sim\mathcal{E}\left(  \sum_{l=1}^{K}\exp(X_{i}%
)^{T}\lambda_{l}\right)  $, we can define EM\ and Gibbs algorithms resp. to
maximize the posterior distribution of the parameters $\lambda_{k}$ and sample
from it when the prior is given by (\ref{eq:prior}).

\section{Experimental results\label{sec:experiments}}

In all the above models, the parameter $b$ is just a scaling parameter on
$\lambda_{k}$. As the likelihood is invariant to a rescaling of the
$\lambda_{k}$, this parameter does not have any influence on inference. Hence
to ensure that the MAP estimate satisfies $\sum_{k=1}^{K}\widehat{\lambda}%
_{k}=1$, we set $b=Ka-1$ henceforth as explained in section
\ref{sec:discussion}. We demonstrate our algorithms on one synthetic and two
real-world data sets.

\subsection{Synthetic Data}

\begin{figure}[h]
\begin{center}
\includegraphics[width=8cm]{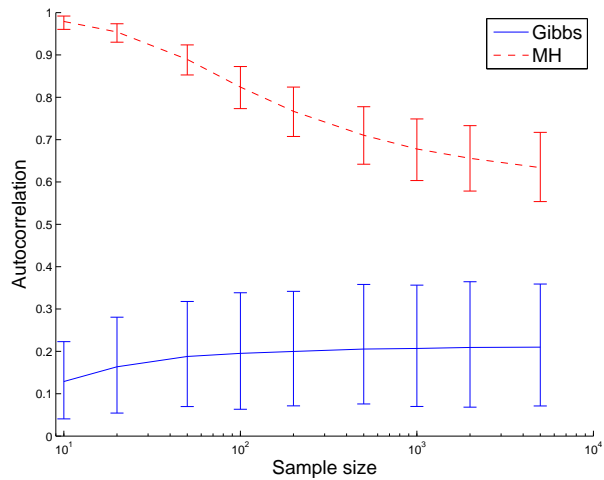}
\end{center}
\caption{Sample lag-1 autocorrelation as a function of the sample size $n$ for
the Gibbs sampler and a modified version of the M-H algorithm of Gormley and
Murphy (2009).}%
\label{fig:ACFsimu}%
\end{figure}

We first study the Plackett-Luce model, comparing experimentally the mixing
properties of the Gibbs sampler relative to a slightly modified version of the
M-H algorithm proposed by~\cite{Gormley2009}. In this latter paper, the
authors propose to update the skill parameters simultaneously using the
following proposal distribution\footnote{The authors actually use a normal
approximation of the gamma distribution, and work with normalized data. To
obtain similar algorithms, we consider unnormalized data.} at iteration $t$
\[
\text{for }i=1,\ldots,K\text{, }\lambda_{i}^{\star}\sim\mathcal{G}\left(
a+w_{k},b+\sum_{i=1}^{n}\left(  \sum_{j=1}^{p_{i}-1}\frac{\delta_{ijk}}%
{\sum_{k=j}^{p}\lambda_{\rho_{ik}}^{(t-1)}}\right)  \right)
\]
We simulated $500$ dataset of $n$ rankings of $K=4$ individuals, for various
values of $n$ with $a=5$. For each dataset, 10,000 iterations of the Gibbs
sampler presented in section \ref{sec:plackettluce} were run. The sample lag-1
autocorrelation was then computed for the four skill parameters. For a given
sample size $n$, the mean value over skill parameters and simulated data is
reported on Figure~\ref{fig:ACFsimu} together with 90\% confidence bounds. The
algorithm of \cite{Gormley2009} performs reasonably well
when the sample size is large, which is the case for the voting data they
considered, but poorly for small sample sizes.

\subsection{Nascar 2002 dataset}

\begin{figure}[h]
\begin{center}
\includegraphics[width=8cm]{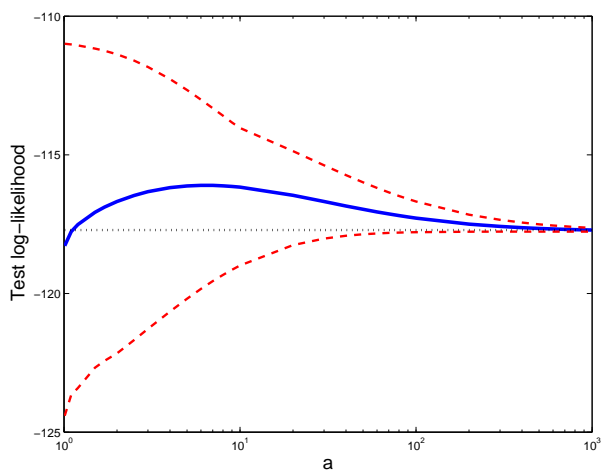}
\end{center}
\caption{Test log-likelihood on the Nascar 2002 dataset. From race 5 to 35, we
compute the log-likelihood of the next race based on the MAP estimates. Mean
and 90\% interval of the log-likelihood is represented w.r.t. to the parameter
$a$. The straight line represents the test log-likelihood obtained with a
uniform prior.}%
\label{fig:nascar2002}%
\end{figure}

NASCAR is the primary sanctioning body for stock car auto racing in the United
States. Each race involves 43 drivers. During the 2002 season, 87 different
drivers participated in 36 races.\ Some drivers participated in all of the
races while others participated in only one. We propose to apply the
Plackett-Luce model with gamma prior on the parameters.\ The NASCAR dataset\footnote{The data can be downloaded from {http://www.stat.psu.edu/~dhunter/code/btmatlab/}}
has been studied by \cite{Hunter2004}, who noted that the MLE\ cannot
be found for the original data set as four drivers placed last in each race
they entered, and therefore had to be removed. This does not need to be done
if we follow a Bayesian approach. We focus here on predicting the outcome of
the next race based on the previous ones, starting from race 5; i.e. we
predict the results of race 6 based on the MAP estimates obtained with the
first 5 races, then the results of race 7 based on the MAP estimated obtained
with the first 6 races, etc. For each race, we compute the test log-likelihood
using the MAP estimates. The mean value and 90\% confidence bounds are
represented in Figure~\ref{fig:nascar2002} w.r.t. the value of $a$. The EM
algorithm was initialized using $(\lambda_{1}^{\left(  0\right)  }%
,\ldots,\lambda_{83}^{\left(  0\right)  })=(\frac{1}{83},\ldots,\frac{1}{83})$.

\begin{figure}[h]
\begin{center}
\includegraphics[width=8cm]{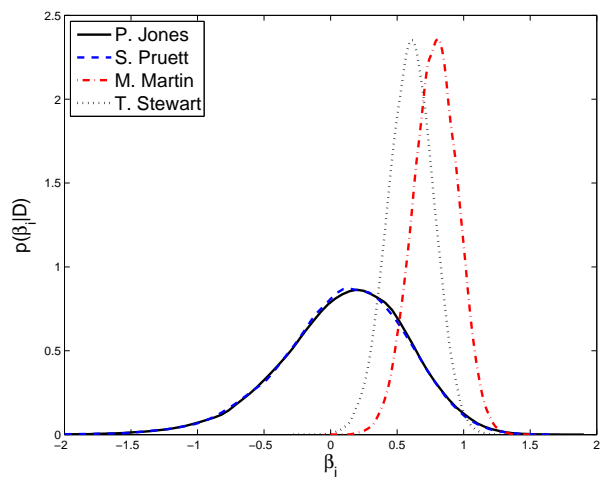}
\end{center}
\caption{Marginal posterior distribution for the modified skill ratings
$\beta_{i}$ of the first 4 drivers according to their average place. P. Jones
and S. Pruett only participated in 1 race each, while M. Martin and T. Stewart
participated in 36 races.}%
\label{fig:postnascar}%
\end{figure}

The Gibbs sampler was also applied to the same dataset. The skill parameters
were initialized at the same value, and the parameter $a$ was assigned a flat
improper prior and sampled as described in section~\ref{sec:discussion}. We
ran 50,000 iterations with 2,000 burn-in. As detailed in
Section~\ref{sec:discussion}, only the normalized weights $\pi_{i}$ are
likelihood identifiable. Skill ratings are usually represented on the real
line, and we use the following one-to-one mapping $\beta_{i}=\log\pi_{i}%
-\log1/83$. The marginal posterior densities of the reparameterized skill
ratings for the first four drivers according to their average place are
reported in Figure~\ref{fig:postnascar}. The Bayesian approach can effectively
capture the uncertainty in the skill ratings of the drivers. ML and MMSE
(minimum mean squared error) estimates together with standard deviations are
reported in Table~\ref{tab:nascar} for the first ten and last ten drivers
according to average place.

\begin{table}[h]
\caption{Top ten and bottom ten drivers according to average place, along with
ML and MMSE estimates of the skill parameters in $\beta$ space. Standard
deviations are also provided.}%
\label{tab:nascar}
\begin{center}%
\begin{tabular}
[c]{cccccc}\hline
&  & Average & MLE & MMSE & Standard\\
Driver & Races & place & estimate & estimate & Deviation\\\hline
P. Jones & 1 & 4.00 & 2.74 & 0.11 & 0.48\\
S. Pruett & 1 & 6.00 & 2.21 & 0.10 & 0.48\\
M. Martin & 36 & 12.17 & 0.67 & 0.79 & 0.17\\
T. Stewart & 36 & 12.61 & 0.42 & 0.60 & 0.17\\
R. Wallace & 36 & 13.17 & 0.65 & 0.78 & 0.17\\
J. Johnson & 36 & 13.50 & 0.53 & 0.68 & 0.17\\
S. Marlin & 29 & 13.86 & 0.33 & 0.49 & 0.19\\
M. Bliss & 1 & 14.00 & 0.82 & 0.04 & 0.48\\
J. Gordon & 36 & 14.06 & 0.33 & 0.53 & 0.17\\
K. Busch & 36 & 14.06 & 0.24 & 0.46 & 0.17\\
... &  &  &  &  & \\
C. Long & 2 & 40.50 & -1.73 & -0.67 & 0.46\\
C. Fittipaldi & 1 & 41.00 & -1.85 & -0.51 & 0.50\\
H. Fukuyama & 2 & 41.00 & -2.17 & -0.81 & 0.50\\
J. Small & 1 & 41.00 & -1.94 & -0.60 & 0.51\\
M. Shepherd & 5 & 41.20 & -1.86 & -1.05 & 0.39\\
K. Shelmerdine & 2 & 41.50 & -1.73 & -0.72 & 0.46\\
A. Cameron & 1 & 42.00 & -1.41 & -0.44 & 0.49\\
D. Marcis & 1 & 42.00 & -1.38 & -0.43 & 0.49\\
D. Trickle & 3 & 42.00 & -1.72 & -0.87 & 0.42\\
J. Varde & 1 & 42.00 & -1.55 & -0.48 & 0.50\\\hline
\end{tabular}
\end{center}
\end{table}

\subsection{Chess data}

Rating the skills of chess players is of major practical interest. It allows
organizers of a tournament to avoid having strong players playing against each
other at early stages, or to restrict the tournament to players with skills
above a given threshold. The international chess federation adopted the
so-called \textquotedblleft Elo\textquotedblright\ system which is based on
the Bradley-Terry model~\citep{Elo1978}.\ For historical considerations about
the rating system in chess, the reader should refer to
\cite{Glickman1995}.

\begin{figure}[h]
\begin{center}
\includegraphics[width=8cm]{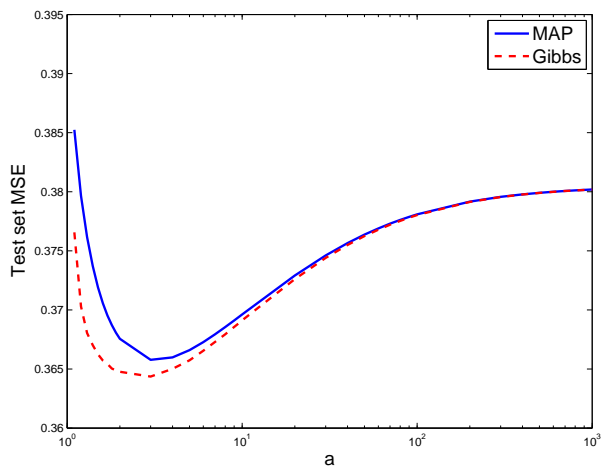}
\end{center}
\caption{Test mean square error on the chess dataset for different values of
the parameter $a$. Based on an history of 95 months, we predict the outcome of
the games of the last 5 months. }%
\label{fig:chesstest}%
\end{figure}

\begin{figure}[h]
\begin{center}
\subfigure[Skill parameter with largest mean value]{\includegraphics[width=6cm]{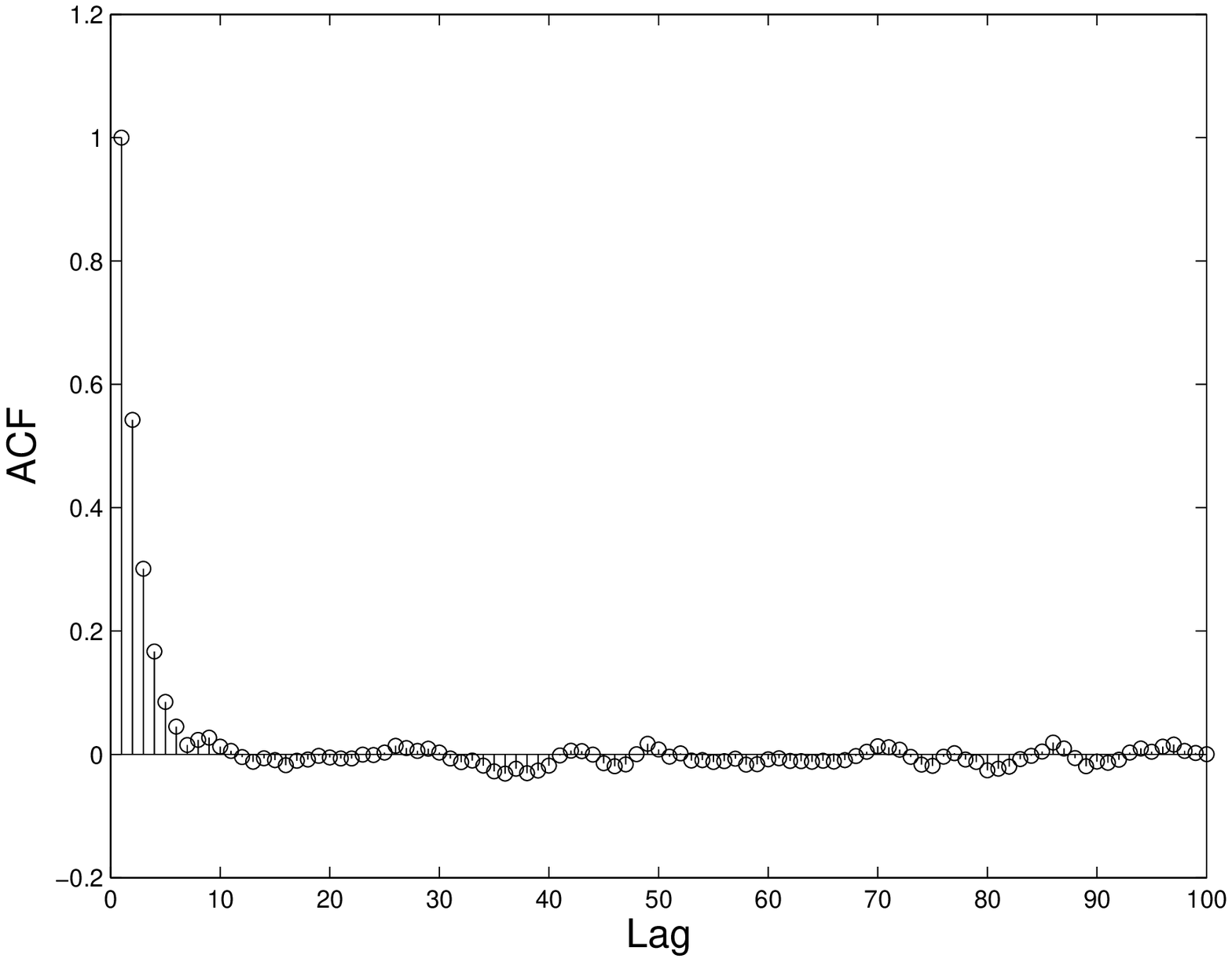}}
\subfigure[Skill parameter with second largest mean value]{\includegraphics[width=6cm]{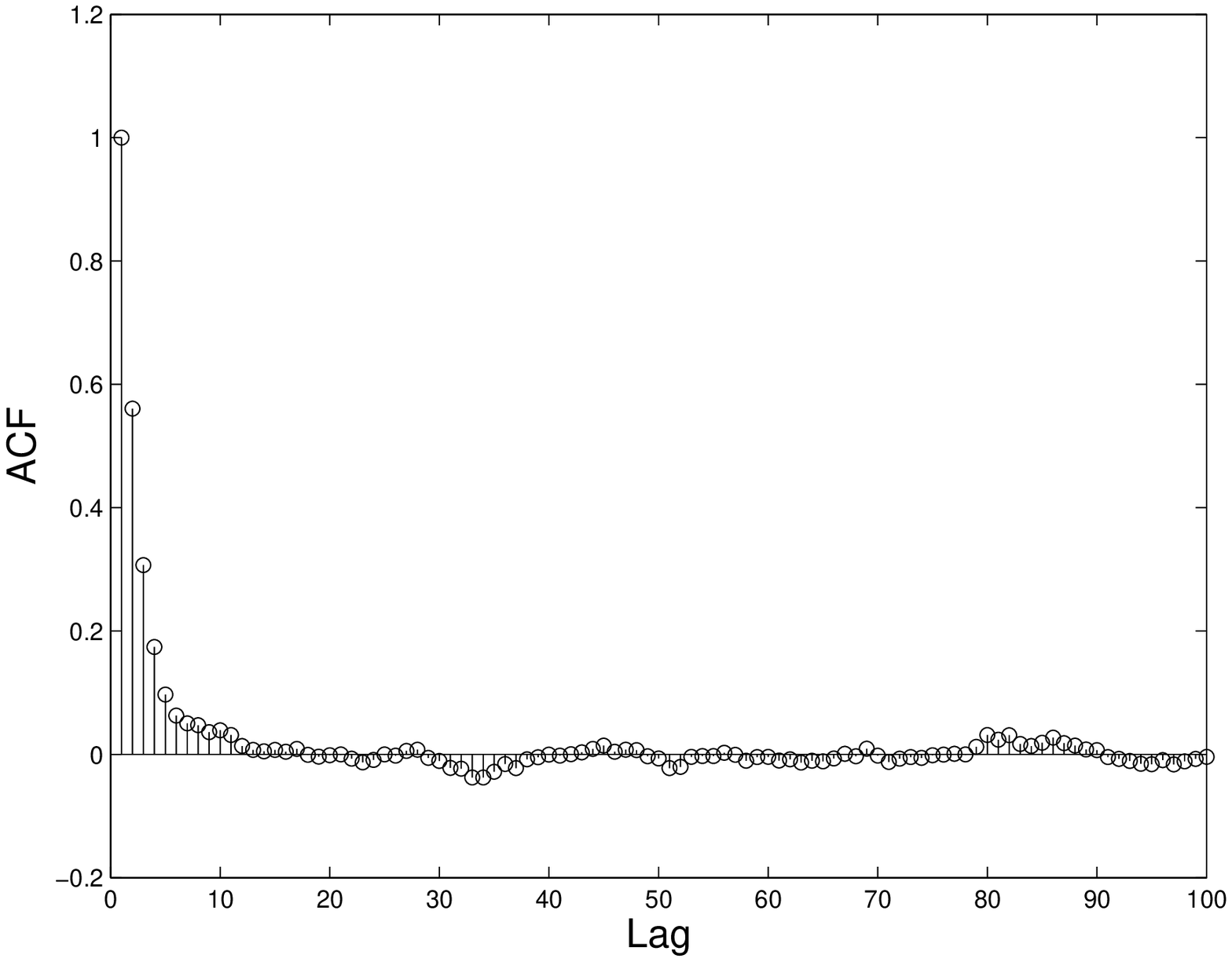}}
\subfigure[Skill parameter with third largest mean value]{\includegraphics[width=6cm]{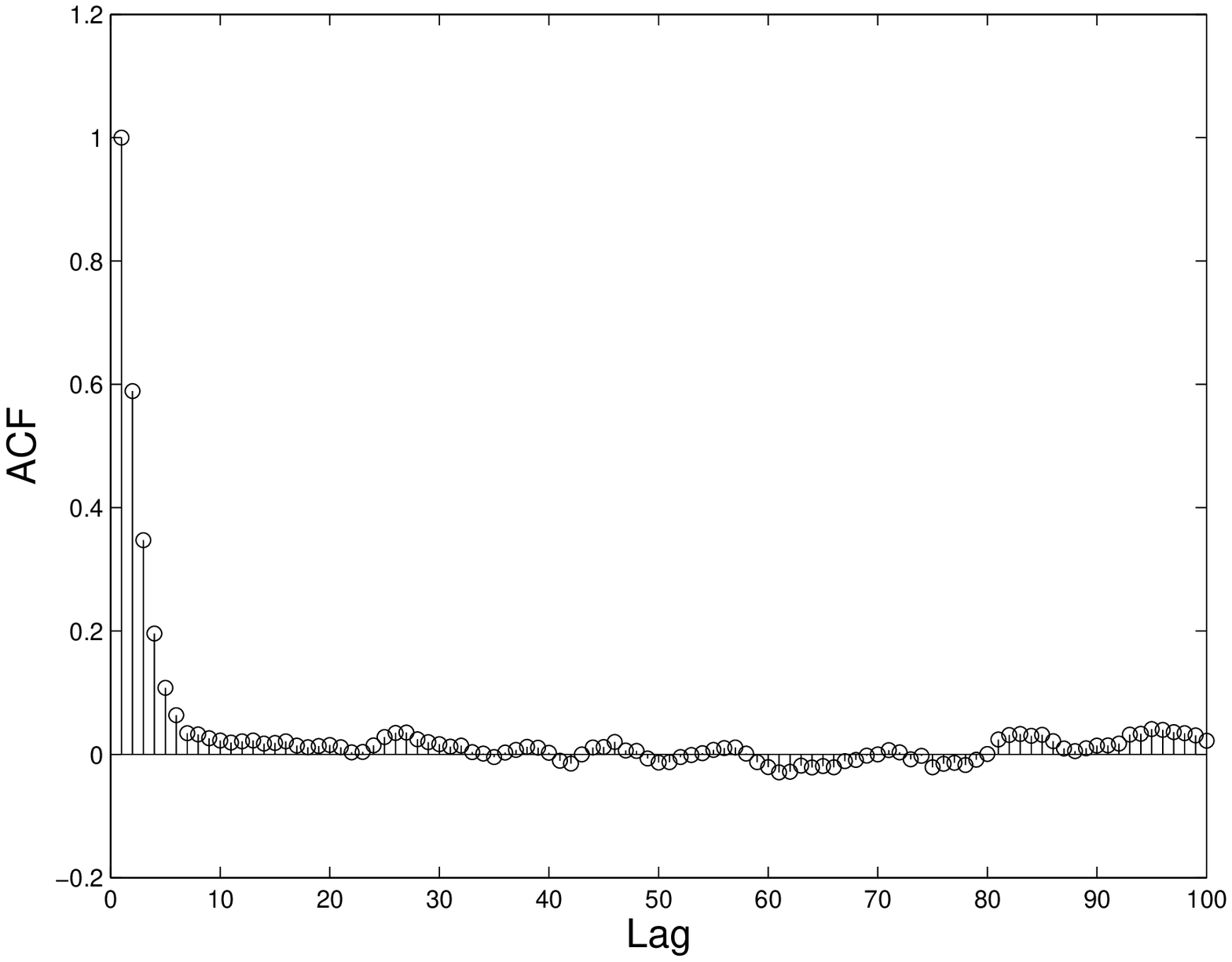}}
\subfigure[Tie parameter $\theta$]{\includegraphics[width=6cm]{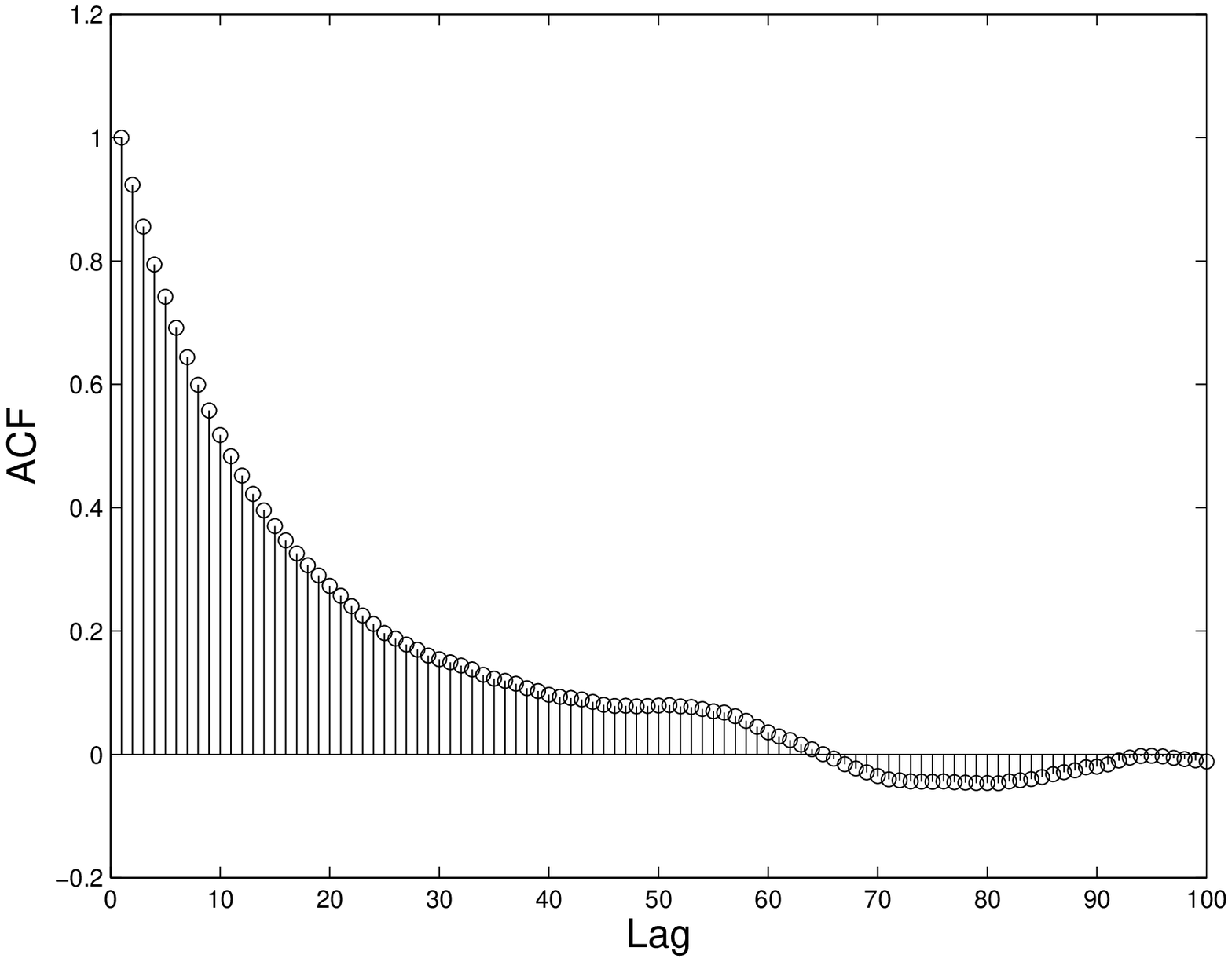}}
\end{center}
\caption{Autocorrelation functions for (a-c) the three skill parameters with
largest mean values and for (d) the parameter $\theta$. The fast decrease
indicates that the Markov chain mixes well. The parameter $\theta$ is updated
with a Metropolis-Hastings step which explains its relatively low mixing.}%
\label{fig:ACF}%
\end{figure}

We consider here game-by-game chess results over 100 months, consisting of
65,053 matches between 8631 players\footnote{Chess data can be downloaded from http://kaggle.com/chess}. The outcome of each game is either win
(+1), tie (+0.5) or loss (0). We estimate the parameters of the Bradley-Terry
model with ties presented in section \ref{sec:ties} on the first 95 months and
then predict the outcome of the games of the last 5 months. The
hyperparameters for the tie parameter $\theta$ are set to $a_{\theta}=1,$
$b_{\theta}=0$. Given the large sample size, it is not possible to sample from
Eq. \eqref{eq:conditionalnonstandardtheta} as the number of elements in the
mixture is very large. We therefore use a M-H step with a normal random walk
proposal of standard deviation $0.1$. The mean squared error is reported for
predictions based on MAP estimates and full\ Bayesian predictive based on the
Gibbs sampler outcomes, for different values of the hyperparameter $a$. EM and
Gibbs samplers were initialized at $(\lambda_{1}^{\left(  0\right)  }%
,\ldots,\lambda_{8631}^{\left(  0\right)  })=(\frac{1}{8631},\ldots,\frac
{1}{8631})$ and $\theta^{\left(  0\right)  }=1,5$. The Gibbs samplers were run
with 10,000 iterations and 1,000 burn-in iterations. The results are reported
in Figure~\ref{fig:chesstest}. The results demonstrate the benefit of
penalizing the skill rating parameters and the improvement brought up by a
full Bayesian analysis. In Figure \ref{fig:ACF} we also report the
autocorrelation function associated to the parameter $\theta$ and the skill
parameters with largest mean values. The Markov chain displays good mixing properties.

\section{Conclusion}

The Bradley-Terry model and its generalizations arise in numerous
applications. We have shown here that most of the MM\ algorithms proposed in
\cite{Hunter2004} can be reinterpreted as special cases of EM algorithms.
Additionally we have proposed original EM\ algorithms for some recent
generalizations of the Bradley-Terry models. Finally we have shown how the
latent variables introduced to derive these EM\ algorithms lead
straightforwardly to Gibbs sampling algorithms. These elegant MCMC\ algorithms
mix experimentally well and outperform a recently proposed M-H algorithm.

\bigskip

{\large Acknowledgment}. The authors are grateful to Persi Diaconis for
helpful discussions and pointers to references on the Plackett-Luce and random
graph models and to Luke Bornn for helpful comments.

\small{
\bibliographystyle{apalike}
\bibliography{ranking}}

\begin{thebibliography}{}

\bibitem[Adams, 2005]{Adams2005}
Adams, E. (2005).
\newblock Bayesian analysis of linear dominance hierarchies.
\newblock {\em Animal Behaviour}, 69:1191--1201.

\bibitem[Agresti, 1990]{Agresti1990}
Agresti, A. (1990).
\newblock {\em Categorical Data Analysis}.
\newblock Wiley.

\bibitem[Barndorff-Nielsen and Shephard, 2001]{Barndorff-Nielsen2001}
Barndorff-Nielsen, O. and Shephard, N. (2001).
\newblock Non-{G}aussian {O}rnstein-{U}hlenbeck-based models and some of their
  uses in financial economics.
\newblock {\em Journal of the Royal Statistical Society B}, 63:167--241.

\bibitem[Bradley and Terry, 1952]{Bradley1952}
Bradley, R. and Terry, M. (1952).
\newblock Rank analysis of incomplete block designs. {I.} the method of paired
  comparisons.
\newblock {\em Biometrika}, 39:324--345.

\bibitem[Chatterjee et~al., 2010]{Chatterjee2010}
Chatterjee, S., Diaconis, P., and Sly, A. (2010).
\newblock Random graphs with a given degree sequence.
\newblock Technical report, Stanford University.

\bibitem[Damien et~al., 1999]{Damien1999}
Damien, P., Wakefield, J., and Walker, S. (1999).
\newblock {Gibbs} sampling for {B}ayesian non-conjugate and hierarchical models
  by using auxiliary variables.
\newblock {\em Journal of the Royal Statistical Society B}, 61:331--344.

\bibitem[David, 1988]{David1988}
David, H. (1988).
\newblock {\em The method of paired comparisons}.
\newblock Oxford University Press.

\bibitem[Davidson and Farquhar, 1976]{Davidson1976}
Davidson, R. and Farquhar, P. (1976).
\newblock A bibliography on the method of paired comparisons.
\newblock {\em Biometrics}, 32:241--252.

\bibitem[Diaconis, 1988]{Diaconis1988}
Diaconis, P. (1988).
\newblock {\em Group representations in probability and statistics, IMS Lecture
  Notes}, volume~11.
\newblock Institute of Mathematical Statistics.

\bibitem[Elo, 1978]{Elo1978}
Elo, A. (1978).
\newblock {\em The rating of Chess Players, Past and present}.
\newblock Arco Pub.

\bibitem[Glickman, 1995]{Glickman1995}
Glickman, M. (1995).
\newblock A comprehensive guide to chess ratings.
\newblock Technical report, Department of Statistics, Boston University.

\bibitem[Gormley and Murphy, 2008]{Gormley2008a}
Gormley, I. and Murphy, T. (2008).
\newblock Exploring voting blocs with the {I}rish electorate: a mixture
  modeling approach.
\newblock {\em Journal of the American Statistical Association},
  103:1014--1027.

\bibitem[Gormley and Murphy, 2009]{Gormley2009}
Gormley, I. and Murphy, T. (2009).
\newblock A grade of membership model for rank data.
\newblock {\em Bayesian Analysis}, 4:265--296.

\bibitem[G\"or\"ur et~al., 2006]{Gorur2006}
G\"or\"ur, D., J\"akel, F., and Rasmussen, C. (2006).
\newblock A choice model with infinitely many latent features.
\newblock In {\em International Conference on Machine Learning}.

\bibitem[Guiver and Snelson, 2009]{Guiver2009}
Guiver, J. and Snelson, E. (2009).
\newblock {B}ayesian inference for {P}lackett-{L}uce ranking models.
\newblock In {\em International Conference on Machine Learning}.

\bibitem[Hastie and Tibshirani, 1998]{Hastie1998}
Hastie, T. and Tibshirani, R. (1998).
\newblock Classification by pairwise coupling.
\newblock {\em Annals of Statistics}, 26:451--471.

\bibitem[Holland and Leinhardt, 1981]{Holland1981}
Holland, P. and Leinhardt, S. (1981).
\newblock An exponential family of probability distributions for directed
  graphs.
\newblock {\em Journal of the American Statistical Association}, 76:33--65.

\bibitem[Huang et~al., 2006]{Huang2006}
Huang, T.-K., Weng, R., and Lin, C.-J. (2006).
\newblock Generalized {B}radley-{T}erry models and multi-class probability
  estimates.
\newblock {\em Journal of Machine Learning Research}, 7:85--115.

\bibitem[Hunter, 2004]{Hunter2004}
Hunter, D. (2004).
\newblock {MM} algorithms for generalized {B}radley-{T}erry models.
\newblock {\em The Annals of Statistics}, 32:384--406.

\bibitem[Lange et~al., 2000]{Lange2000}
Lange, K., Hunter, D., and Yang, I. (2000).
\newblock Optimization transfer using surrogate objective functions (with
  discussion).
\newblock {\em Journal of Computational and Graphical Statistics}, 9:1--59.

\bibitem[Liu, 2001]{Liu2001}
Liu, J. (2001).
\newblock {\em Monte Carlo Methods for Scientific Computing}.
\newblock Springer-Verlag: New York.

\bibitem[Luce, 1959]{Luce1959}
Luce, R. (1959).
\newblock {\em Individual choice behavior: A theoretical analysis}.
\newblock Wiley.

\bibitem[Luce, 1977]{Luce1977}
Luce, R. (1977).
\newblock The choice axiom after twenty years.
\newblock {\em Journal of Mathematical Psychology}, 15:215--233.

\bibitem[Park and Newman, 2004]{Park2004}
Park, J. and Newman, M. (2004).
\newblock The statistical mechanics of networks.
\newblock {\em Physical Review E}, 70:066117.

\bibitem[Plackett, 1975]{Plackett1975}
Plackett, R. (1975).
\newblock The analysis of permutations.
\newblock {\em Applied Statistics}, 24:193--202.

\bibitem[Rao and Kupper, 1967]{Rao1967}
Rao, P. and Kupper, L. (1967).
\newblock Ties in paired-comparison experiments: A generalization of the
  {B}radley-{T}erry model.
\newblock {\em Journal of the American Statistical Association}, 62:194--204.

\bibitem[Restle, 1961]{Restle1961}
Restle, F. (1961).
\newblock {\em Psychology of judgement and choice}.
\newblock New-York: Wiley.

\bibitem[Tversky, 1972a]{Tversky1972}
Tversky, A. (1972a).
\newblock Choice by elimination.
\newblock {\em Journal of Mathematical Psychology}, 9:341--367.

\bibitem[Tversky, 1972b]{Tversky1972a}
Tversky, A. (1972b).
\newblock Elimination by aspects: a theory of choice.
\newblock {\em Psychological Review}, 79:281--299.

\bibitem[Wickelmaier and Shmidt, 2004]{Wickelmaier2004}
Wickelmaier, F. and Shmidt, C. (2004).
\newblock A {M}atlab function to estimate choice model parameters from
  paired-comparison data.
\newblock {\em Behavior Research Methods, Instruments and Computers},
  36:29--40.

\bibitem[Zermelo, 1929]{Zermelo1929}
Zermelo, E. (1929).
\newblock Die berechnung der turnier-ergebnisse als ein maximumproblem der
  wahrscheinlichkeitsrechnung.
\newblock {\em Math. Z.}, 29:436--460.

\end{thebibliography}

\end{document}